\documentclass[a4paper,12pt]{article}

\usepackage{amsmath}
\usepackage{amsfonts}

\begin{document}
\title{
\begin{small}
\hspace{11.8cm}
\begin{tabular}{r}
UWThPh-2000-41\\
October 2000
\end{tabular}
\end{small}
\\
\vspace{1cm}
A generalized Bell-Inequality
and Decoherence
for the $K^0\bar K^0$-System}

\author{
Beatrix C. Hiesmayr\\
{\em Institute for Theoretical Physics, University of Vienna}\\
{\em Boltzmanngasse 5, A-1090 Vienna, Austria}\\
\date{October 9, 2000}}
\maketitle
\begin{abstract}
First a generalized Bell-inequality for different times and for different quasi-spin
states is developed. We focus on special quasi-spin eigenstates and times. The inequality
based on a
local realistic theory is violated by the $CP$-violating parameter \cite{Uchiyama},
if the quantum theory is used to recalculate the probabilities.
Next the quantum mechanical probabilities are modified by the decoherence approach
which enables the initial state to factorize spontaneously. In this way we get a lower limit
for the decoherence parameter $\zeta$, which measures the degree of decoherence.
This result is compared with the experimental value \cite{Article1, trixi} of the
decoherence parameter $\zeta$  deduced from the data of the
CPLEAR-experiment \cite{CPLEAR-EPR}.
\end{abstract}
\vspace{2cm}

\section{Introduction}
Tests of quantum mechanics (QM) against local realistic hidden
variable theories are of great interest since the first formulation
of the EPR paradox by Einstein, Podolski
and Rosen in 1935 \cite{EPR}. J.S. Bell \cite{bell} proved in 1964
the important theorem that a whole class of local realistic hidden
variable theories cannot reproduce all statistical
predictions of QM.

Another approach was done by Wigner. He started from a pure
set-theoretical point of view, where he - simply spoken - counted all
super-pure states, respectively, all possible measurable results of such states.

However,
to test such inequalities experimentally, a better approach is
the Bell-CHSH-inequality (Clauser-Horne-Shimony-Holt)
\cite{bell3, CHSH, Clauser}, because it doesn't require perfect anti-correlation
nor perfect measurement apparatus. These experiments using correlated photons
have been done by many groups \cite{aspect, zeilinger} and the results agree with the
quantum mechanical predictions.

In recent years people started to consider
massive EPR-like correlated particle systems, see for instance
Ref.\cite{lee} to Ref.\cite{SelleriBook}.
One of them is the neutral kaon system, where a
to the photon case similar singlet-state can be produced.

In this work we are going to develop a Bell-CHSH-inequality based on a hidden variable theory
for the neutral entangled kaon
system where both the
times and the quasi-spin eigenstates can be differed. This turns out to be a
generalization, which allows us to handle existing Bell-inequalities derived with different
methods in an uniform way.

The variation of the times when a special quasi-spin
eigenstate is measured is analogous to the spin-case, however, the
theoretical calculations show that a Bell-CHSH-inequality
cannot be
violated by quantum mechanics due to the specific constants
of that neutral kaon system \cite{ghirardi91}. If the quantity $x=\frac{2 \Delta m}{\gamma_S}$
was
a factor $4.3$ smaller than the experimental $x \approx 1$, the Bell-CHSH-inequality would
be violated \cite{trixi}.

But if one allows to vary the quasi spin eigenstate the Bell-CHSH-inequality can be violated by
quantum mechanics. Choosing three
special eigenstates, namely $|K_S\rangle$, $|\bar K^0\rangle$ and $|K_1^0\rangle$, one can
transform the Bell-CHSH-inequality into a Wigner-type
inequality. Then by inserting the quantum mechanical
joint probabilities into this Wigner-type inequality one gets an
inequality on the $CP$-parameter $\varepsilon$ \cite{Uchiyama}. This Wigner-type
inequality is violated by the experimental value of the $CP$-parameter
$\varepsilon$ \cite{Particle Data}.

We want to bring to the readers attention the importance of using the correct time
evolution for the derivation of the quantum mechanical probabilities, see section
\ref{quantummechanicalprobability}. Our Bell-CHSH-inequality differs from the inequality
in literature, e.g., \cite{domenico}, we obtain an additional correlation function,
since we deal with a unitary time evolution.

Next we consider another method, the decoherence method, which has been developed
for the massive neutral correlated kaon-system
in \cite{Article1, trixi, eberhard} and for the massive correlated
neutral B-mesons
\cite{BG}. This method modifies quantum mechanics in the way that the quantum
mechanical interference term is multiplied by
by a factor $(1-\zeta)$, where the decoherence
parameter $\zeta$ equal $0$ refers to QM and $\zeta$
equal $1$ refers to a vanishing interference term, this case is called total
decoherence or Furry's hypothesis\footnote{This hypothesis, actually, should be called Schr\"odinger's
hypothesis, first because he published such a factorization of an entangled state one
year \cite{Schrödinger} before W. Furry \cite{Furry} and second he stated that such a
happening really could occur.} or spontaneous factorization. In the neutral kaon system the value of the decoherence parameter $\zeta$
has been calculated using the data of the CPLEAR-experiment \cite{CPLEAR-EPR} at CERN,
the results are published in \cite{Article1}.

In order to find a closer connection between the
decoherence approach and local deterministic theories we use quantum probabilities
derived
with that simple manipulation to recalculate the Wigner-type inequality.

\section{The Bell-CHSH-inequality for the K-mesons}

Here we will derive a Bell-CHSH-inequality based on locality, realism and induction
for different times and for arbitrary quasi spin states in the massive EPR-correlated neutral
kaon system. First we will focus on the similarities and differences of the photon system compared
with the neutral kaon system.

\subsection{Introductory considerations}

Testing the predictions of quantum mechanics against those of any local
deterministic hidden variable theory presents some analogies but also significant
practical and conceptual differences with respect to the corresponding problem in
the case of spin variables. The differences derive from two specific features.

\begin{enumerate}
\item First, while in the spin or photon case one can devise a test to check whether a
spin $\frac{1}{2}$-particle is or is not in any chosen spin state $\alpha |\Uparrow_n\rangle+
\beta |\Downarrow_n\rangle$, there is no analogous way to test
whether the system is in the linear superposition
$\alpha |K^0\rangle+\beta |\bar K^0\rangle$. However, as done in \cite{Uchiyama,
domenico} we will assume that the four following superpositions of the strangeness
eigenstates can be measured by a gedanken experimentator

\begin{eqnarray}
\textrm{Mass-eigenstates:}& &|K_S\rangle=\frac{1}{N}\big\lbrace p |K^0\rangle-q
|\bar K^0\rangle \big\rbrace\nonumber\\
& &|K_L\rangle=\frac{1}{N}\big\lbrace p |K^0\rangle+q
|\bar K^0\rangle \big\rbrace\nonumber\\
CP\textrm{-eigenstates:}& &|K_1^0\rangle=\frac{1}{\sqrt{2}}\big\lbrace |K^0\rangle-
|\bar K^0\rangle \big\rbrace\nonumber\\
& &|K_2^0\rangle=\frac{1}{\sqrt{2}}\big\lbrace |K^0\rangle+
|\bar K^0\rangle \big\rbrace
\end{eqnarray}

with $p=1+\varepsilon$, $q=1-\varepsilon$, $N^2=|p|^2+|q|^2$ and $\varepsilon$ being
the complex $CP$-violating parameter. So far, the following derivations and conclusions
belong to a gedanken experiment, but in the end we just need the information of the
$CP$-violating parameter which can be measured by an arbitrary experiment not necessarily dealing with
entangled particles.

\item The second important difference drives from the fact that while in the
spin-case the direct product space $H^1_{spin} \otimes H^2_{spin}$ is sufficient to account for
all spin properties of the entangled system, this is \emph{not} true for the
neutral kaon case.

As a consequence the norm of the component on such a space decreases with time, if one doesn't
take the decay products into account; that is the main difference between the works
of, e. g., \cite{domenico} and \cite{ghirardi91}.

Indeed, we want to emphasize that in the case under consideration the state vector
acquires
by time evolution components on the
manifold of the decay products which are orthogonal to the product space $H^r_{kaon} \otimes
H^l_{kaon}$. Then we get a unitary time evolution, which leads to additional terms
in the resulting Wigner-type inequality, see section \ref{thegeneralCHSH-inequality}.
\end{enumerate}

\subsection{Requirements of locality}

In  the case of spin variables one can derive the
Bell-CHSH-inequality \cite{bell3, CHSH}, for the averaged values of spin directions along
arbitrary quantization directions $n$ and $m$. The analogue of the free choice of the
spin directions is, in the kaon case, the free choice of the times at which
measurements aimed to detect the quasi spin states of the meson. But, in addition, we have
the freedom of choosing the
quasi spin state of the meson, the strangeness-eigenstate, the mass-eigenstate or
the $CP$-eigenstate.

The locality assumption requires then that the results at one side be completely
independent of the time and the choice of the quasi spin eigenstates at which
the measurement at the other side is performed. To define the appropriate
correlation functions to be used in Bell's inequality, one considers an observable
$O^r(k_n,t_a)$ on the right side, which assumes the value $+1$ if the measurement at
time point $t_a$ gives the quasi spin eigenstate $k_n$ and the value $-1$
if the quasi spin eigenstate $k_n$ is not found. In terms of such an observable we can define the correlation
function $O(k_n t_a; k_m t_b)$, which takes the value $+1$ both when a $k_n$ at $t_a$
and a $k_m$ at $t_b$ was detected or when no $k_n$ and no $k_m$ was detected. In the case that
only one of the desired quasi spin eigenstate has been found, no matter at which
side, the correlation function takes the value $-1$.

The locality assumption implies then that $O(k_n t_a, k_m t_b)$, in a specific
individual experiment, equals the product of $O^r(k_n, t_a)$ and $O^l(k_m, t_b)$:

\begin{eqnarray}
O(k_n t_a; k_m t_b)&=& O^r(k_n, t_a)\;\cdot\; O^l(k_m, t_b).
\end{eqnarray}

From this equation taking one derives immediately

\begin{eqnarray}
|O(k_n t_a; k_m t_b)- O(k_n t_a; k_{m'} t_c)|\;+\;|O(k_{n'} t_d; k_{m'} t_c)+O(k_{n'} t_d;
k_m t_b)|\;= 2
\end{eqnarray}

with $k_n, k_m, k_{m'}$ and $k_{n'}$ being arbitrary quasi spin eigenstates of the meson
and $t_a, t_b, t_c$ and $t_d$ four different times.

Let us now consider a sequence of $N$ identical measurements, and let us denote by $O_n$
the value taken by $O$ in the $i$-th experiment. The average is then given by

\begin{eqnarray}
M(k_n t_a; k_m t_b)&=&\frac{1}{N} \sum_{i=1}^N O_i(k_n t_a; k_m t_b)
\end{eqnarray}

and satisfies the Bell-CHSH-inequality \cite{bell3, CHSH}

\begin{eqnarray}\label{chsh-inequality}
& &| M(k_n t_a; k_m t_b)-M(k_n t_a; k_{m'} t_c)|+| M(k_{n'} t_d; k_{m'} t_c)+
M(k_{n'} t_d; k_m t_b)|\; \leq \nonumber\\
& &\frac{1}{N} \sum_{i=1}^N \big\lbrace
|O_i(k_n t_a; k_m t_b)- O_i(k_n t_a; k_{m'} t_c)|\;+\;|O_i(k_{n'} t_d; k_{m'} t_c)+O_i(k_{n'} t_d;
k_m t_b)|\;=\;2.\nonumber\\
\end{eqnarray}

\section{How to derive the quantum probabilities?}\label{quantummechanicalprobability}

We will follow here the formalism described in \cite{ghirardi91}, but generalize it
using both different times and arbitrary quasi spin-states.
The complete evolution of the mass-eigenstates is described by a unitary
operator $U(t,0)$ whose effect can be written as

\begin{eqnarray}\label{zeitentwicklung}
U(t,0)\; |K_{S,L}\rangle&=& e^{-i \lambda_{S,L} t}\;|K_{S,L}\rangle +
|\Omega_{S,L}(t)\rangle
\end{eqnarray}

where $|\Omega_{S,L}(t)\rangle$ describes the decay products. Thus we operate in a
complete Hilbert space in opposite to, i.e., \cite{domenico}.

The initial state of the strong decay of the $\Phi$-meson, $J^{PC}=1^{--}$ into a
pair of neutral K-mesons is given in the $K^0 \bar K^0$-basis and $K_S K_L$-basis choice by

\begin{eqnarray}\label{initialstate}
& &K^0 \bar K^0\textrm{-basis:}\;\;|\psi(t=0)\rangle=\frac{1}{\sqrt{2}}\big\lbrace |K^0\rangle
|\bar K^0\rangle-|\bar K^0\rangle |K^0 \rangle\big\rbrace\nonumber\\
& &K_S K_L\textrm{-basis:}\;\;|\psi(t=0)\rangle=\frac{N^2}{2 p q \sqrt{2}}\big\lbrace |K_S\rangle
|K_L\rangle-|K_L\rangle |K_S \rangle\big\rbrace.
\end{eqnarray}

The state at time $t$ is then obtained from (\ref{initialstate}) by applying to it a
unitary operator which is the direct product

\begin{eqnarray}\label{zeitaufspaltung}
U(t,0)&=& U_r(t,0)\cdot U_l(t,0)
\end{eqnarray}

of the operators $U_r(t,0)$ and $U_l(t,0)$ acting on the space of the right and the
left mesons in accordance with (\ref{zeitentwicklung}).

According to standard quantum mechanics we will now evaluate the probabilities of
finding different quasi spin eigenstates in measurements at two different times $t_r$ and
$t_l$; without loss  of generality $t_r>t_l$. We denote by $P_r(k_n)$ the
projection operator on the right side projecting the quasi spin-eigenstate $k_n$, so
that, e. g. $P_r(K^0)=|K^0\rangle_r \langle K^0|_r$. As usual the projection
operator $Q_{r}(k_n)=1-P_{r}(k_n)$ acts on the manifolds orthogonal to those
associated to $P_{r}(k_n)$.

Starting from the initial state (\ref{initialstate}) one gets  with
(\ref{zeitaufspaltung})
at time $t_l$ the state

\begin{eqnarray}
|\psi(t_l,t_l)\rangle&=& U(t_l,0)|\psi(t=0)\rangle\;=\;U_r(t_l,0) U_l(t_l,0)
|\psi(t=0)\rangle.
\end{eqnarray}

If we now measure a $k_m$ on the left side at $t_l$ this state yields reduction to
the state

\begin{eqnarray}
|\tilde{\psi}(t_l, t_l)\rangle&=&P_l(k_m) |\psi(t_l, t_l)\rangle.
\end{eqnarray}

Now we need to evaluate this state to the time $t_r$ and project on the right state

\begin{eqnarray}\label{uu1}
|\tilde{\psi}(t_r, t_l)\rangle&=& P_r(k_n)U_r(t_r,t_l)
P_l(k_m)|\psi(t_l,t_l)\rangle.
\end{eqnarray}

The state of that equation (\ref{uu1}) gives the probability of finding a mesons
in the state $k_n$ on the right side at time $t_r$ and a $k_m$ state at the left
side at time $t_l$. Such a state, taking into account the unitarity and the
composition laws of the operator $U$ as well as the fact that operators referring to
the different (right and left) Hilbert spaces commute, coincides with the state

\begin{eqnarray}\label{uu2}
|\psi(t_r,t_l)\rangle&=& P_r(k_n) P_l(k_m) U_r(t_r,0) U_l(t_l,0) |\psi(t=0)\rangle.
\end{eqnarray}

In this publication we will consider probabilities of finding or not finding a specific quasi
spin state in a specific measurement. Derivations of the corresponding probabilities for such
a  process can be done by using formula (\ref{uu2}) with the operators $Q$ replacing
the operators $P$, where required.

For example, the joint probabilities that in two measurements at $t_r$ and $t_l$ a quasi-spin
state $k_n$ is detected (Y) and a quasi-spin state $k_m$ is not detected (N) is
given by

\begin{eqnarray}
P_{n,m}(Y t_r,N t_l)&=&||P_r(k_n) Q_l(k_m) U_r(t_r,0) U_l(t_l,0)
|\psi(t=0)\rangle||^2.
\end{eqnarray}

\section{The general Bell-CHSH-inequality}\label{thegeneralCHSH-inequality}

The derivation of the quantum mechanical probabilities for finding on the right side
at $t_a$ a quasi-spin state $k_n$ or not and on the left side at $t_b$ a quasi-spin
state $k_m$ or not has been shown in the last section. Hence, we can write the quantum mechanical expectation value for
finding an arbitrary state $k_n$ at time $t_a$ on the right side and a state $k_m$ at time $t_b$
on the left side

\begin{eqnarray}\label{qmmeanvalue}
& &M^{QM}(k_n t_a, k_m t_b)=\nonumber\\
& &P_{n,m}(Y t_a, Y t_b)+P_{n,m}(N t_a, N t_b)-P_{n,m}(Y t_a, N t_b)
-P_{n,m}(N t_a, Y t_b)\nonumber\\
\end{eqnarray}

where $P_{n,m}(Y t_a, Y t_b)$ is the probability of finding a $k_n$ at $t_a$ on the right
side and finding a $k_m$ at $t_b$ on the left side; $P_{n,m}(Y t_a, N t_b)$ denotes the
case when we find a $k_n$ at $t_a$, but our detector doesn't detect a $k_m$ at
$t_b$.

Further we can use that the sum of the statistical frequencies of the results $(Y,Y)$,
$(N,N)$, $(Y,N)$ and $(N,Y)$ is one for all times, so Eq.(\ref{qmmeanvalue}) can be
rewritten to

\begin{eqnarray}
M^{QM}(k_n t_a, k_m t_b)=-1 + 2 \big\lbrace P_{n,m}(Y t_a, Y t_b)+P_{n,m}(N t_a,N t_b) \big\rbrace.
\end{eqnarray}

Setting this expression into the Bell-CHSH-inequality (\ref{chsh-inequality}) we get the
following inequality

\begin{eqnarray}\label{diegleichung}
& &|P_{n,m}(Y t_a, Y t_b)+P_{n,m}(N t_a, N t_b)-P_{n,m'}(Y t_a, Y t_c)-P_{n,m'}(N t_a, N t_c)|\; \leq \;\nonumber\\
& &\qquad \qquad
1\pm\big\{-1+1\;\{\hphantom{+}\;P_{n',m}(Y t_d, Y t_b)+P_{n',m}(N t_d, N t_b)\nonumber\\
& &\hphantom{\qquad \qquad
1\pm\big\{-1+1\;\{}
+P_{n',m'}(Y t_d, Y t_c)+ P_{n',m'}(N t_d, N t_c)\}\big\}.
\end{eqnarray}

\subsection{The Wigner-type inequality}

To derive from this Bell-CHSH-inequality (\ref{diegleichung}) the Wigner-type inequality
we have to choose the upper sign $+$ and we get

\begin{eqnarray}\label{allgemeinewigner}
P_{n, m}(Y t_a,Y t_b)&\;\leq\;&P_{n,m'}(Y t_a,Y t_c)+P_{n',m}(Y t_d,Y t_b)+P_{n',m'}(Y t_d,Y t_c)\nonumber\\
& &+
h(n,m,n',m';t_a,t_b,t_c,t_d)
\end{eqnarray}

with
\begin{eqnarray}\label{correctionfunctions}
h(n,m,n',m';t_a,t_b,t_c,t_d)&=&-P_{n,m}(N t_a,N t_b)+P_{n,m'}(N t_a,N t_c)+
P_{n',m}(N t_d,N t_b)\nonumber\\
& &+P_{n',m'}(N t_d,N t_c)
\end{eqnarray}

Setting $t_a=t_b=t_c=t_d=t=0$ the function $h(n,m,n',m';t=0)$ is equal to

\begin{eqnarray}
h(n,m,n',m';t=0)&=&-P_{n,m}(Y,Y)|_{t=0}+P_{n,m'}(Y,Y)|_{t=0}+P_{n',m}(Y,Y)|_{t=0}\nonumber\\
& &+P_{n',m'}(Y,Y)|_{t=0}
\end{eqnarray}

because $P(Y,Y)|_{t=0} \equiv P(N,N)|_{t=0}$.
To get rid of the fourth probability we use the anti-correlation of the entangled
system, setting $n'$ equal to $m'$ this probability becomes zero. Thus we derive the following
Wigner-type inequality at $t=0$

\begin{center}
\begin{tabular}{|c|}
\hline
$\vphantom{\biggr\rbrace}P_{n,m}(Y,Y)|_{t=0}\;\leq\;P_{n,n'}(Y,Y)|_{t=0}+P_{n',m}(Y,Y)|_{t=0}$\\
\hline
\end{tabular}
\end{center}

This inequality was found by Uchiyama \cite{Uchiyama} by a set-theoretical approach.
He showed that for choosing

\begin{eqnarray}\label{choice}
& &|k_n\rangle\; = |K_S\rangle\nonumber\\
& &|k_m\rangle = |\bar K^0\rangle\nonumber\\
& &|k_{n'}\rangle = |K_1\rangle
\end{eqnarray}

the Wigner-type inequality is violated by the $CP$-violating
parameter $\varepsilon$:

\begin{eqnarray}\label{Wigner}
Re\{\varepsilon\}\;\leq\;|\varepsilon|^2.
\end{eqnarray}

Higher orders in the $CP$-violating parameter of $\varepsilon$ are neglected. This
Wigner-type inequality (\ref{Wigner}) is obviously violated by the experimental value of
$\varepsilon$, having an absolute value of about $10^{-3}$ and a phase of about $45^\circ$.

If we would replace the anti-kaon with a kaon, thus choose our three states in the
following way

\begin{eqnarray}
& &|k_n\rangle\; = |K_S\rangle\nonumber\\
& &|k_m\rangle = |K^0\rangle\nonumber\\
& &|k_{n'}\rangle = |K_1\rangle
\end{eqnarray}

we end with the inequality

\begin{eqnarray}
-Re\{\varepsilon\}\;\leq\;|\varepsilon|^2.
\end{eqnarray}

which is obviously not violated.

On the other hand replacing the short living state $|K_S\rangle$ by the long living
state $|K_L\rangle$ and the CP-eigenstate $|K_1\rangle$ by $K_2\rangle$ we find the same
inequality as (\ref{Wigner}).

Note, when deriving the Wigner-type inequality (\ref{Wigner}) from a set theoretical approach
the
orthogonality of the mass-eigenstates is not needed; the states are only mutually orthogonal
of the
order of $O(|\varepsilon|)$ but not of the order $O(|\varepsilon|^2)$. In this
gedanken experiment, just one mass-eigenstate is sufficient, one
could reformulate the approach by assuming to the mass-eigenstate $|K_S\rangle$ an
orthogonal one, analogously to \cite{domenico}.

However, our aim is not to measure the states $|K_S\rangle$ and $|K_1^0\rangle$. The
Wigner-type inequality for the choice (\ref{choice}) is not physical in the sense
of experimentally testable,
but using the quantum mechanical formalism we get an inequality (\ref{Wigner}) for a physical
quantity, the $CP$-parameter $\varepsilon$. This result will be connected to a
modified theory in section
\ref{theinequalityparameterizedbythedecoherenceparameter}.

\subsection{The Wigner-type inequality for equal times}\label{thewignertypeinequalityfortimesgreaterzero}

Now we will consider the Wigner-type inequality of the previous section for times greater
zero, but equal-time measurements. We have to
pay attention to the
correction function $h(n,m,n',m';t)$.

For the choice (\ref{choice})
the correction function $h(n,m,n',m';t)$ is given by

\begin{eqnarray}
& &h(K_S, \bar K^0, K^0_1, K^0_1; t)=2+\frac{1}{1-x^2} \big\lbrace
\frac{-2}{1+|\varepsilon|^2} e^{-\gamma_S t}
+\frac{-2|\varepsilon |^2}{1+|\varepsilon|^2} e^{-\gamma_L t}\nonumber\\
& &\hphantom{f_1(K_S, \bar K^0, K^0_1, K^0_1; t)=}+
(1-x^2) \frac{|\varepsilon|^2-Re\{\varepsilon\}}{1+|\varepsilon|^2} e^{-2 \gamma t}
+2 x^2 \cos(\Delta m t)\cdot e^{-\gamma t}\big\rbrace\nonumber\\
\end{eqnarray}

It turns out that the time-dependent Wigner-type inequality is only violated for
times smaller than $t=8\cdot10^{-4} \tau_S$, hence,
for times larger this value the Wigner inequality is restored.

\subsection{The Wigner-type inequality for different times}

To avoid the fast increase of the correction function $h$ the times can be chosen
$t_a=t_c=t_d$ with $t_a \leq t_b$. The violation of the Wigner-type inequality
is strongest for $t_a$ close to zero; in this case a violation for $t_b$ up to $4\tau_S$
can be found.

\section{The Wigner-type inequality parameterized by the decoherence parameter $\zeta$}\label{theinequalityparameterizedbythedecoherenceparameter}

Now we want to bring the decoherence parameter $\zeta$ into the game. We are going
to recalculate the probabilities needed for the Wigner inequality
(\ref{allgemeinewigner}), but with possible decoherence, i.e., the interference term is
multiplied by the factor $(1-\zeta)$ where $\zeta$ is the decoherence parameter,
already used in \cite{Article1, trixi, BG, eberhard}. This simple modification
gives the neutral kaon
system the possibility of spontaneous factorization of the initial state (\ref{initialstate}).

The idea here is to determine which degree of decoherence, i.e., which value of the
decoherence parameter $\zeta$ is sufficiently large, to restore the inequality
(\ref{allgemeinewigner}). In this way we can relate the decoherence approach to a local
realistic theory.

This procedure is
basis dependent, thus it depends on which basis the spontaneous factorization occurs in,
i.e. in which basis
the cross products terms are affected by the decoherence. In the publications
\cite{Article1, trixi} it has been demonstrated that
using the
experimental data of the CPLEAR experiment at CERN \cite{CPLEAR-EPR} the value of
the decoherence parameter $\zeta$ can be calculated.

Further it has been demonstrated that there are two physical interesting basis choices, the $K_S
K_L$- and the $K^0 \bar K^0$-basis choice. We will start with the $K_S K_L$-basis
choice and continue with the $K^0 \bar K^0$-basis.

\subsection{Calculations in the $K_S K_L$-basis}

Denoting with $k_n$ and $k_m$ as usual the quasi-spin states the joint probability
$P_{n,m}(Y t_r,Y t_l)$ can be
derived
according to QM - starting from the initial anti-symmetric state in the $K_S
K_L$-basis (\ref{initialstate}) and using the formalism of section
\ref{quantummechanicalprobability} - to

\begin{eqnarray}
& &P_{n,m}^\zeta(Y t_r, Y t_l)\;=\,||P_r(k_n) P_l(k_m) U_r(t_r) U_l(t_l) \frac{N^2}{2 p q
\sqrt{2}}
\big\lbrace |K_S\rangle_r |K_L\rangle_l-|K_L\rangle_r |K_S\rangle_l\big\rbrace||^2\nonumber\\
&=&\frac{N^4}{8 |p|^2 |q|^2}\cdot\nonumber\\
& &||\langle k_n|U_r(t_r)|K_S\rangle\;
\langle k_m|U_l(t_l)|K_L\rangle\;\;|k_n\rangle_r |k_m\rangle_l\nonumber\\
& &\hphantom{||\langle k_n|U_r(t_r)|K_S\rangle\;
\langle k_m|U_l(t_l)|K_L\rangle\;\;}-
\langle k_n|U_r(t_r)|K_L\rangle\;
\langle k_m|U_l(t_l)|K_S\rangle\;\;|k_n\rangle_r |k_m\rangle_l||^2\nonumber\\
&\equiv&\frac{N^4}{8 |p|^2 |q|^2}\cdot\biggl\lbrace\nonumber\\
& &|\langle k_n|U_r(t_r)|K_S\rangle\;
\langle k_m|U_l(t_l)|K_L\rangle|^2+|\langle k_n|U_r(t_r)|K_L\rangle\;
\langle k_m|U_l(t_l)|K_S\rangle|^2\nonumber\\
& &-2\underbrace{(1-\zeta)} Re\{\langle k_n|U_r(t_r)|K_S\rangle^*\;
\langle k_m|U_l(t_l)|K_L\rangle^*\;\langle k_n|U_r(t_r)|K_L\rangle\;
\langle k_m|U_l(t_l)|K_S\rangle\}\biggr\rbrace\nonumber\\
& &\textrm{Modification}
\end{eqnarray}

with $U_{r,l}(t_{r,l}) \equiv U_{r,l}(t_{r,l}, 0)$.
However, if we calculate the joint probability $P_{n,m}^\zeta(Y t_r, N t_l)$ it gets more
complicating

\begin{eqnarray}
& &P_{n,m}^\zeta(Y t_r, N t_l)\;=\,||P_r(k_n) Q_l(k_m) U_r(t_r) U_l(t_l) \frac{N^2}{2 p q
\sqrt{2}}
\big\lbrace |K_S\rangle_r |K_L\rangle_l-|K_L\rangle_r |K_S\rangle_l\big\rbrace||^2\nonumber\\
&=&\frac{N^4}{8 |p|^2 |q|^2}\cdot\nonumber\\
& &||\biggl(\langle k_n|U_r(t_r)|K_S\rangle\;\;|k_n\rangle_r U_l(t_l)|K_L\rangle_l
-\langle k_n|U_r(t_r)|K_S\rangle\;\langle k_m|U_l(t_l)|K_L\rangle\;\;|k_n\rangle_r |k_m\rangle_l\biggr)\nonumber\\
& &-
\biggl(\langle k_n|U_r(t_r)|K_L\rangle\;\;|k_n\rangle_r U_l(t_l)|K_S\rangle_l
-\langle k_n|U_r(t_r)|K_L\rangle\;\langle k_m|U_l(t_l)|K_S\rangle\;\;|k_n\rangle_r |k_m\rangle_l\biggr)||^2\nonumber\\
&\equiv&\frac{N^4}{8 |p|^2 |q|^2}\cdot\biggl\lbrace\nonumber\\
& &|\biggl(\dots\biggr)|^2+|\biggl(\dots\biggr)|^2\nonumber\\
& &-2\underbrace{(1-\zeta)} Re\{\biggl(\dots\biggr)^*\biggl(\dots\biggr)\}\biggr\rbrace.\nonumber\\
& &\textrm{Modification}
\end{eqnarray}

Deriving the $\zeta$-terms for all probabilities under consideration and setting $t=0$ the
correction
function $h^{K_S K_L}_\zeta$ (\ref{correctionfunctions}) which includes all $(N,N)$ probabilities
gives

\begin{eqnarray}
\lefteqn{h^{K_S K_L}_\zeta(K_S, \bar K^0, K^0_1, K^0_1; t=0)=}\nonumber\\
& &\qquad\qquad\frac{-Re\{\varepsilon\}+|\varepsilon|^2}{2(1+|\varepsilon|^2)}+\frac{\zeta}{1-x^2} \biggl\lbrace
\frac{x(1+x)}{4}+\frac{|\varepsilon|^2}{(1+|\varepsilon|^2)^2}\biggr\rbrace\nonumber\\
\end{eqnarray}

with $x=\frac{2 Re\{\varepsilon\}}{1+|\varepsilon|^2}$. Note, in opposite to pure quantum mechanics
for this modified
quantum theory the probability $P^{\zeta}_{n,n}(Y,Y)|_{t=0}$ and $P^{\zeta}_{n,n}(N,N)|_{t=0}$
is not necessarily zero for $\zeta\neq0$.

For further simplification we will approximate the $CP$-violating parameter $\varepsilon$
by setting the phase
equal $45^\circ$ and the parameters\footnote{This is the superweak-model, which assumes
$\varepsilon'$ to be 0 \cite{Wolfenstein}. However $\varepsilon'$ not equal zero
wouldn't effect our results.} $\eta_{00}=\eta_{+-}=\varepsilon$:

\begin{eqnarray}
& &Re\{\varepsilon\}\;\approx\; Im\{\varepsilon\}\geq 0\nonumber\\
& &\qquad\Longrightarrow |\varepsilon|^2\approx 2 Re^2\{\varepsilon\}:=2u^2.
\end{eqnarray}

Thus the approximate numerical value of $|\varepsilon|$ is $(2.28\pm0.019)\cdot10^{-3}$
with a phase of $45^\circ$ \cite{Wolfenstein}.

The calculation of all with the decoherence parameter modified probabilities of inequality
(\ref{allgemeinewigner}) gives the following
inequality for the decoherence parameter $\zeta$ to the order $O(u^2)$

\begin{eqnarray}\label{ksklresult}
\frac{u-2 u^2}{u+6 u^2}=0.987\leq \zeta.
\end{eqnarray}

This is our important result, it is of great interest because to satisfy this inequality the
decoherence-parameter $\zeta$ has to be very close to 1; or in other words the
interference term has to vanish. Hence, Furry's hypothesis or spontaneous
factorization has to take place totally. This means in our case that after the
creation of the entangled pair the initial state vector (\ref{initialstate}) factorizes
in 50\% of the cases in a short living state at the right side and in a long living
state at the left side or in the other 50\% of the cases vice versa.

Intuitively, we would have expected that there exist local realistic theories which
allow at least partially an interference term, see for instance \cite{six2,SelleriBook}.
Our result demands for a vanishing interference term, hence, the
locality assumptions underlying this inequality forces the $K_S K_L$-interference term to
vanish.

On the other hand this result can be compared with the results published in \cite{Article1,trixi}. In
these publications a CPLEAR-experiment at CERN \cite{CPLEAR-EPR} in 1998 was considered. They
produced entangled kaons and measured the strangeness-states.
The average result for the decoherence parameter is
$\zeta^{K_S K_L}=0.13\lbrace^{0.16}_{0.15}$. Within one standard deviation quantum
mechanics $\zeta=0$ is included, but total decoherence $\zeta=1$ is excluded by many
standard deviations.

This means for experimental reasons that $\zeta=1$ is excluded and due to the
Bell-inequality (\ref{ksklresult}) a local realistic variable theory is impossible.

\subsection{Calculations in the $K^0 \bar K^0$-basis}

As we have seen in \cite{Article1} the $K^0\bar K^0$-basis choice is not a good one
to discriminate between the quantum theory and Furry's hypothesis or spontaneous
factorization. Let's see what we get if we recalculate the Wigner-type inequality ($t=0$)
this time
starting by the $K^0 \bar K^0$-basis choice.

After a cumbersome calculation one gets for $t=0$

\begin{eqnarray}\label{kkresult}
u(1-2 u^2)=0.0016 \; \leq \; \zeta.
\end{eqnarray}

Hence, if $\zeta$ is just a little bigger than the small value $10^{-3}$, the
Wigner-type
inequality will be restored and a local realistic theory which obeys the assumptions
which lead to that
Wigner-type inequality will be possible.

Again this result can be compared with the decoherence parameter for the $K^0 \bar
K^0$-basis choice derived with the help of the CPLEAR-data, which is
$\zeta^{K^0 \bar K^0}\sim 0.4\pm0.7$.
In this basis choice the initial state (\ref{initialstate}) can factorize in a kaon at
the right side and an anti-kaon at the left side or vice versa. Hence, within one
standard deviation the quantum mechanical result $\zeta=0$ is included, but also
total decoherence $\zeta=1$ is within one standard deviation.

On the other hand it should be possible to construct in that basis choice  a local
realistic theory  which obeys the locality assumption of the derived inequality with
a $K^0 \bar K^0$-interference term which differs just by the order of $O(u)$
from the quantum mechanical result.

\section{Conclusion}

We develop a Bell-CHSH-inequality and simplify it to a Wigner-type inequality. Inserting the quantum
mechanical probabilities into that inequality we find for $t=0$ a Wigner-type inequality,
the inequality of F. Uchiyama \cite{Uchiyama},
which is violated by the value of the $CP$-violating parameter; thus we have a
contradiction between a local hidden variable theory and the prediction of the
quantum theory.

Next we consider a modified quantum theory, which describes possible spontaneous
factorization of the initial state, where the measure of decoherence $\zeta$ has two
limits, the quantum theory $\zeta=0$ on one hand and a local theory $\zeta=1$ on the
other hand. Inserting such a modified quantum theory into a Bell-CHSH-inequality results in
an inequality for the decoherence parameter $\zeta$.

The simple modification was calculated for two
basis choices, the $K_S K_L$-basis choice and the $K^0 \bar K^0$-basis choice. The
result for the first choice was that the decoherence parameter $\zeta$ had to be very
close to $1$ to fulfill the Wigner-type inequality and thus obey the locality
assumption. Intuitively, we would have expected that there exist local
realistic theories which allow at least partially an interference
term in that basis choice, see for instance \cite{six2, SelleriBook}.

However, comparing with the range of the decoherence parameter \cite{Article1}
derived with the data of the CPLEAR-experiment \cite{CPLEAR-EPR} we can exclude
$\zeta=1$.

In the $K^0 \bar K^0$-basis choice the situation is different.
The derived Wigner-type inequality (\ref{kkresult}) is satisfied
for all $\zeta$s being larger than the small value of about
$10^{-3}$. Comparing that result with the range of the
decoherence parameter $\zeta$
derived in \cite{Article1}, we learn that this basis
choice doesn't give us - in contrast to the `best' basis choice, $K_S K_L$ -
a powerful tool to distinguish between a realistic
local theory obeying the Wigner-type inequality and QM on the
other side.

Concluding, we have shown that there exists a best basis choice
where
one can distinguish clearly first to which extent decoherence could take
place and second if within that range a local theory is possible.
Thus with connecting the two different approaches we can exclude a local realistic
theory.\\
\\

\textbf{ACKNOWLEDGMENT}

I specially want to thank R.A. Bertlmann and W. Grimus for
enlightening and encouraging discussions and suggestions.
This work is partly supported by Austrian FWF project P14143-PHY, the Magistratsabteilung  18 - Stadtentwicklung und
Stadtplanung (MA 18-W/758/99) and the Austrian-Czech Republic
Scientific Collaboration, project KONTAKT 1999-8.

\end{document}